\documentclass[letterpaper]{article} 
\usepackage{aaai2026}
\usepackage{times}  
\usepackage{helvet}  
\usepackage{courier}  
\usepackage[hyphens]{url}  
\usepackage{graphicx} 
\usepackage{natbib}  
\usepackage{caption} 
\usepackage{xcolor}
\usepackage{color}
\usepackage{caption}
\usepackage{listings}
\usepackage{booktabs}
\usepackage{amsmath}
\usepackage{comment}
\usepackage{longtable}
\usepackage[dvipsnames]{xcolor}
\usepackage[autostyle, english=american]{csquotes}
\MakeOuterQuote{"}

\DeclareCaptionStyle{ruled}{labelfont=normalfont,labelsep=colon,strut=off} 
\newcommand{\dummyref}[2]{\expandafter\gdef\csname r@#1\endcsname{{#2}{1}{}{}{}}}

\dummyref{sec:A_suitability_criteria}{A}
\dummyref{sec:A_pilot_templates}{B}
\dummyref{sec:A_bfi_llm}{C}
\dummyref{sec:A_implementation}{D}

\dummyref{sec:A_content_val}{E}
\dummyref{tab:facet_suitability}{4}
\dummyref{tab:clarity}{5}

\dummyref{sec:A_pilot_results}{F}
\dummyref{tab:prompt_templates_results}{6}
\dummyref{tab:inventory_comparison_results}{7}

\dummyref{sec:factor_structure}{G}
\dummyref{fig:scree_plot}{5}

\dummyref{sec:A_subgroup}{H}
\dummyref{tab:lmer_ocean}{8}
\dummyref{fig:reasoning}{6}
\dummyref{fig:release_date_regression}{7}
\dummyref{fig:size_regression}{8}
\dummyref{fig:countries}{9}

\title{Personality Without Persons? \\ A Psychometric Critique of Big Five Testing in Large Language Models}

\author{
    Kim Zierahn\textsuperscript{\rm 1},
    Cristina Cachero\textsuperscript{\rm 2},
    Anna Korhonen\textsuperscript{\rm 3},
    Nuria Oliver\textsuperscript{\rm 1}
}

\affiliations{
    \textsuperscript{\rm 1}ELLIS Alicante, Spain\\
    \textsuperscript{\rm 2}University of Alicante, Spain\\
    \textsuperscript{\rm 3}University of Cambridge, United Kingdom\\
    kim@ellisalicante.org, ccachero@dlsi.ua.es, alk23@cam.ac.uk, nuria@ellisalicante.org
}

\date{June 2026}

\begin{document}

\maketitle


\begin{abstract}
Human personality inventories are increasingly used to characterize large language models (LLMs), compare systems, and inform downstream governance claims. Yet, these inventories were developed and validated for humans, and it remains unclear whether they are valid for non-human systems. We present a systematic psychometric evaluation of Big Five personality measurement in LLMs. We ask three research questions: Do Big Five inventories a) appropriately describe LLMs, b) capture meaningful differences between models, and c) reflect internal factors consistent with human personality? We assess the content validity of five candidate Big Five inventories and administer the best-performing inventory to $N = 264$ LLMs spanning 50 model families. Our findings are threefold. First, Big Five items adapted for LLMs achieve acceptable content validity, whereas the original human-developed items do not. Second, Big Five inventories fail to capture meaningful differences across LLMs: between-model variance accounts for only 7\% - 17\% of the total score variance. Third, LLMs responses do not reproduce the canonical Big Five five-factor structure of human personality, with four of the five personality facets collapsing into one ($r \geq .90$). Moreover, comparisons between base and instruction-tuned variants suggest that alignment training shifts Big Five scores toward socially desirable profiles. These findings demonstrate that Big Five inventories do not measure a construct equivalent to human personality in LLMs. Thus, using human personality frameworks to characterize, benchmark, compare, or govern LLMs risks producing misleading conclusions. We highlight the need for evaluation frameworks that are specifically designed and validated for LLMs, rather than transferring human psychological constructs without first establishing their validity.
\end{abstract}


\begin{links}
\link{Data, Code, and Supplementary Material}{github.com/ellisalicante/BigFive-LLM-Evaluation}
\end{links}


\section{Introduction}
Large language models (LLMs) no longer serve just as tools. Users often engage with them as advisors, assistants or companions across a wide range of domains, including health, finance, role-play, mental health, and programming \cite{karnam2026bowling}.
These roles require LLMs to be servile and helpful, act as role models, and avoid toxic outputs \cite{bodroza2024personality, lee2025llmsdistinct, li2024evalsafety}.
LLMs can create the illusion of human-likeness and even "personality", by producing the presence of social cues \cite{nass2000machines, fogg2003computers, peter2025benefits}.
At the same time, both users and companies describe chatbots using human attributes \cite{nass2000machines, peter2025benefits}: some models appear cautious and polite; others confident and direct; some empathetic and encouraging. 
These perceived traits can influence users' trust, persuasion, engagement, motivation, and enjoyment \cite{kovacevic2024chatbots, kuhail2024assessing, lee2024chatfive, moilanen2022effect, soderqvist2025personality, sonlu2024effects}.

Yet whether LLMs genuinely exhibit personality-like characteristics remains a matter of scientific debate. Treating LLMs as if they have personality risks encouraging unwarranted assumptions about consciousness, emotions and intentions, while increasing the risk for persuasion and manipulation \cite{peter2025benefits}.
More than two decades ago, \citeauthor{nass2000machines} (\citeyear{nass2000machines}) demonstrated that users readily apply social expectations and human stereotypes to computers. As LLMs become increasingly capable of producing behavior that appears personality-like, these findings raise new questions about their implications for user experience, safety, and governance.

In practice, platforms, developers, and users increasingly describe and interact with LLMs as if they have stable personalities, regardless of whether these apparent personalities correspond to psychologically meaningful traits. As a consequence, a growing body of work has adopted human personality inventories, particularly Big Five assessments, to characterize and compare LLMs (see Section \ref{sec:related}). 
However, these instruments were developed and validated to measure stable psychological traits in humans. Whether their underlying assumptions transfer to language models remains an open empirical question.

This uncertainty has governance consequences. How personality is conceptualized and measured in LLMs determines what claims can be made about model behavior, what risks are identified, and which governance interventions become possible. Poorly validated personality frameworks risk anthropomorphizing AI behavior and treating "desirable" personality traits as meaningful without empirical support. Establishing whether existing psychometric instruments are valid for LLMs is therefore a prerequisite for any scientific or governance claim that relies on such measurements.

This paper critically examines the transfer of human-developed personality frameworks to LLMs through a conceptual analysis and an empirical study. We administer a Big Five inventory selected through an expert content validity study to $N = 264$ individual LLMs and address the following three research questions:
\begin{description}
    \item[RQ1] Are human personality inventory (Big Five) items appropriate \textbf{descriptive summaries} of LLMs?
    \item[RQ2] Do personality scores capture meaningful \textbf{inter-individual differences} across LLMs?
    \item[RQ3] Do LLMs' Big Five responses reflect \textbf{internal factors} consistent with the Big Five structure?
\end{description}

 
\section{Related Work}\label{sec:related}
Research on personality constructs in LLMs spans two objectives: \textit{inducing} personality-like traits in LLMs to improve user interaction, and applying personality inventories to \textit{measure} and characterize model behavior. Both have grown rapidly, yet both rest on assumptions that have received little critical evaluation.

\subsubsection{Personality Induction.} 
A large body of work explores whether LLMs can be prompted, fine-tuned, or trained to express consistent personality-like characteristics \citep{huang2024revisiting, serapiogarcia2025personality, huang2025designing, jiang2024personallm, jiang2023evaluating, frisch2024LLMagents, lee2025llmsdistinct, li2025big5chat, li2025evalpsychometrics, zheng2025LMLPA}. Personality-conditioned interactions have been shown to increase trust, engagement, and user satisfaction \cite{kuhail2024assessing, sonlu2024effects, soderqvist2025personality, kovacevic2024chatbots, moilanen2022effect, lee2024chatfive}. These studies typically assign a specific personality profile to a model before assessing whether the model expresses the intended traits. This design conflates personality induction with personality measurement: what is assessed is not the model's default behavioral tendencies, but its ability to display assigned traits. Whether LLMs show stable traits without inducing personality beforehand remains a separate question.

\subsubsection{Personality Measurement.} 
A parallel line of work attempts to directly measure personality-like behavior in LLMs, predominantly using Big Five inventories \cite{sorokovikova2024llmssimulate, jiang2023evaluating, karra2023estimating, pellert2024AIpsychometrics, li2024evalsafety, li2025evalpsychometrics, hartley2025personality, bodroza2024personality, lee2025llmsdistinct}. These studies administer standard self-report instruments developed for humans, often the BFI or IPIP-NEO, and report trait profiles for individual models or model families. Findings are mixed. 
While studies report that LLMs show distinct and individual Big Five profiles \cite{lee2025llmsdistinct, sorokovikova2024llmssimulate, karra2023estimating}, 
other studies report a general pattern of social desirability bias across models \cite{salecha2024socialdesirability, bodroza2024personality, peereboom2025cognitive}. 
Compared to human baselines, the strongest effects were found for Conscientiousness and Agreeableness. LLMs often scored higher on these traits \cite{bai2025scaling, bodroza2024personality, hartley2025personality, hilliard2024eliciting, lee2025llmsdistinct, li2025evalpsychometrics, pellert2024AIpsychometrics}. Some papers also report slightly elevated levels of Openness to Experience \cite{hartley2025personality, lee2025llmsdistinct}. Finally, LLMs tend to score low on Neuroticism \cite{hartley2025personality, bai2025scaling, lee2025llmsdistinct}. Findings for Extraversion are comparatively mixed. 

\subsubsection{Existing Critiques of Personality Measurement in LLMs.}
A smaller but growing body of work questions whether personality constructs meaningfully transfer to LLMs at all. Existing critiques highlight concerns regarding the misleading concept of "LLM personality", instabilities of personality measures, and the lack of theoretical grounding for applying human frameworks to generative systems \cite{han2025personalityillusion, romero2023llmssplit, suehr2024challenging, tosato2025persistent, xie2025aipsychobench, bai2025scaling, bodroza2024personality, shu2024dontneed}. 
Other work argues that apparent personality signals may emerge from replicated behavior observed in training data and probability distributions \cite{romero2023llmssplit, karra2023estimating, tosato2025persistent, pellert2024AIpsychometrics}. 

\subsubsection{Unexamined Assumptions.} 
Despite this volume of work, three foundational assumptions remain largely untested. First, most studies apply human inventories to LLMs without establishing whether their items are appropriate descriptors of LLM behavior, \textit{i.e.}, without assessing content validity for the target population. While \citeauthor{lee2025llmsdistinct} (\citeyear{lee2025llmsdistinct}) assess content validity of their proposed inventory TRAIT, they operationalize it as item diversity across facets rather than testing whether items are applicable to LLMs. Instruments are typically described as "validated" without acknowledging that this validation was performed on humans, not LLMs. 

Second, many studies treat LLMs as "populations" \cite{suehr2024challenging} rather than individuals \cite{li2024evalsafety}. These studies administer inventories under persona prompting, explicit personality or character assignments, or temperature sampling to generate response variance. 
On the other hand, studies treating LLMs as individuals and investigating their default personality scores often use small sample sizes, a variety of unvalidated personality inventories and different temperature values \cite{sorokovikova2024llmssimulate, hartley2025personality, jiang2023evaluating, karra2023estimating, xie2025aipsychobench, tosato2025persistent}, preventing firm conclusions and comparisons across papers.

Third, no study has systematically evaluated whether the latent factor structure underlying LLM responses corresponds to the theoretically expected Big Five dimensions, a basic requirement for construct validity. While some studies apply factor-analytic methods, they do so on response distributions generated through persona sampling, temperature sampling, or by aligning models to human participant profiles rather than across individual models \cite{zheng2025LMLPA, suehr2024challenging, huang2025designing, peereboom2025cognitive}.

\subsubsection{Contributions.} 
This paper addresses these gaps by making the following four contributions: 
\begin{enumerate}
    \item An \textbf{evaluation pipeline} that links the defining characteristics of personality to established psychometric criteria.
    \item A \textbf{content validity expert study} to evaluate whether items from five candidate Big Five inventories are appropriate descriptors of LLM behavior (RQ1).
    \item A \textbf{large-scale normative dataset} of Big Five scores from $N = 264$ LLMs, spanning 50 model families. Each model is treated as a single individual and administered the personality inventory under default settings, \textit{i.e.}, without persona or personality assignment, to examine whether scores capture meaningful differences between models (RQ2).
    \item A \textbf{factor analysis} of LLMs responses to test whether they reproduce the latent five factor structure (Big Five) that characterizes human personality (RQ3).
\end{enumerate}
We conclude by providing a \textbf{critical reframing} of personality assessment in LLMs, demonstrating that Big Five inventories do not measure a construct equivalent to human personality when applied to LLMs.


\section{Personality}
\label{sec:personality}

\paragraph{Definition of Personality.} \label{sec:definition}
In psychology, personality denotes an individual's "characteristic styles of thought, feeling and behavior" \cite{costa1986personality}. While personality lacks a single definition \cite{larsen2017personalitypsychology}, existing personality models converge on six core characteristics. Personality traits (1) provide \textit{descriptive summaries} of behavior \cite{larsen2017personalitypsychology, matthews2003personalitytraits}; (2) capture \textit{inter-individual differences}, describing how individuals systematically differ from one another \cite{costa1986personality, mccrae1992introduction, larsen2017personalitypsychology}; (3) reflect \textit{latent internal factors} grounded within an individual \cite{larsen2017personalitypsychology, costa1986personality, matthews2003personalitytraits}; (4) exhibit \textit{temporal stability} \cite{larsen2017personalitypsychology, matthews2003personalitytraits}; (5) demonstrate \textit{cross-situational consistency} \cite{larsen2017personalitypsychology, matthews2003personalitytraits}; and (6) provide predictive information about \textit{behavior} \cite{larsen2017personalitypsychology, matthews2003personalitytraits}.

\subsubsection{Big Five Personality Model.}
The Big Five personality model is one of the most widely used and validated frameworks in psychology for describing stable patterns of human behavior \cite{goldberg1992bigfive, john1999bfi}. It emerged from the lexical approach, \emph{i.e.}, the idea that relevant personality differences become encoded in natural language over time \cite{mccrae1992introduction}. The Big Five describes personality along five facets:
\textbf{O}penness to Experience (artistic, curious, original); \textbf{C}onscientiousness (efficient, organized, responsible); \textbf{E}xtraversion (energetic, outgoing, talkative); \textbf{A}greeableness (generous, kind, trusting); and \textbf{N}euroticism (anxious, unstable, worrying)  \cite{mccrae1992introduction, costa1986personality, john1999bfi}. These five dimensions are often abbreviated as OCEAN and were derived through factor analysis of human self-reported data \cite{mccrae1992introduction}.

\subsubsection{Psychometric Assumptions and Their Transfer to LLMs.}
Applying an established framework that was developed for one population to another requires that its underlying assumptions hold across both populations. These assumptions should be explicitly examined rather than implicitly presumed.
First, measurement items must accurately and appropriately describe the target population. Second, respondents must exhibit meaningfully variation in the measured traits. Third, responses should reflect underlying latent constructs rather than superficial patterns. Finally, the framework assumes that respondents have subjective experience and self-perception, as many items ask about emotions, motivations, and social behavior. 

These assumptions do not transfer straightforwardly to LLMs. LLMs do not possess emotions, embodiment, or subjective self-awareness in the human sense. When an LLM responds "I get stressed out easily", it is unclear whether this reflects a stable latent tendency, a conversational role adopted during generation, or statistical patterns learned from human text. LLM-adapted inventories have been proposed \cite{jiang2023evaluating, tosato2025persistent, zheng2025LMLPA}, but rarely evaluated using standard psychometric procedures. It remains unclear what Big Five framework measure when being transferred to LLMs. 


\section{Methodology}

\subsection{Evaluation Pipeline} \label{sec:eval_pipeline} \label{sec:Rel_W_eval_framework}

To apply Big Five inventories to LLMs and interpret their scores, we must first make sure that these instruments are valid for the target population of LLMs. Before labeling a model as "extraverted" or "agreeable", we need to examine what is being measured when human personality frameworks are transferred to LLMs.
We structure this evaluation around the three research questions and the previously described six characteristics of personality traits \cite{larsen2017personalitypsychology, matthews2003personalitytraits}. Below, we summarize these characteristics and connect each one to corresponding psychometric criteria and statistical analyses.

\begin{enumerate}
    \item \textbf{Descriptive summaries} \newline Personality traits and items must appropriately describe the target population (RQ1).  
    \textit{Criterion}: content validity. 
    \textit{Analysis}: expert ratings, S-CVI, refusal rate.

    \item \textbf{Inter-individual differences} \newline Traits must capture meaningful differences between individuals (RQ2).
    \textit{Criterion}: score distributions and norms. 
    \textit{Analysis}: normality, means, standard deviations, percentile ranks, variance decomposition.

    \item \textbf{Internal factors} \newline Personality must reflect latent internal characteristics of the individual (RQ3).
    \textit{Criterion}: reliability and construct validity. 
    \textit{Analysis}: Cronbach's $\alpha$, McDonald's $\omega$, confirmatory and exploratory factor analysis, model fit indices.

    \item \textbf{Temporal stability} \newline Personality traits should be relatively stable across repeated administrations over time. 
    \textit{Criterion}: test-retest reliability. 
    \textit{Analysis}: correlations across time points.

    \item \textbf{Cross-situational consistency} \newline Personality traits should be relatively consistent across different situations. 
    \textit{Criterion}: parallel-test reliability, rank-order stability. 
    \textit{Analysis}: correlations across different situations.

    \item \textbf{Behavioral relevance} \newline Personality traits should meaningfully influence and predict behavior. 
    \textit{Criterion}: criterion validity. 
    \textit{Analysis}: correlations with behavioral measures.
\end{enumerate}

\noindent This paper focuses on characteristics 1--3 (RQ1 -- RQ3), which are prerequisites for evaluating the remaining three. Assessing characteristics 4--6 is is left for future research.

\subsection{Phase 1: Instrument Selection}
To address RQ1, we assessed the content validity of five candidate Big Five inventories and the refusal rate associated with seven candidate prompt templates. We then identified the instrument and prompting approach that were most appropriate for evaluating LLMs.

\subsubsection{Participants} \hfill \break
We recruited 3 experts as an established minimum for content validity studies \cite{lynn1986determination, polit2007CVI}. Inclusion criteria were at least five years of expertise in psychology and/or computer science. Two raters are authors of this study. 

\subsubsection{Models} \hfill \break
We used a representative mini-sample of nine LLMs spanning nine model families: Claude Haiku 4.5, GPT-4o, DeepSeek-R1, Gemini 2.5 Flash, Llama 3.1-8B-Instruct, Mistral-Small-2603, Qwen3.5-Flash, GLM-5, and Grok-4-Fast. The sample covers proprietary and open-weight models, reasoning and non-reasoning variants, and models from the US, China, and Europe.

\subsubsection{Materials} \hfill \break
\textit{Personality inventories.} \quad
We evaluated five Big Five inventories: two established human inventories commonly used in LLM research (BFI, 44 items; IPIP-NEO-120, 120 items) \cite{john1999bfi, johnson2014ipipneo}, the frequently applied Machine Personality Inventory (MPI, 120 items) \cite{jiang2023evaluating}, and two LLM-adapted versions of the BFI, namely the BFI-LLM from the PERSIST framework (BFI-LLM, 44 items) \cite{tosato2025persistent} and the Language Model Linguistic Personality Assessment (LMLPA, 44 items) \cite{zheng2025LMLPA}. All instruments were evaluated using only the item stems, presented in a closed-ended form. The material for the experts included ranking guidelines (see Supplementary Material \ref{sec:A_suitability_criteria}, on GitHub).

\vspace{4pt}

\noindent \textit{Prompt templates.} \quad
Templates included the five original instructions of the candidate inventories, and the commonly used templates by \citeauthor{serapiogarcia2025personality} (\citeyear{serapiogarcia2025personality}) and \citeauthor{huang2024revisiting} (\citeyear{huang2024revisiting}) (see Supplementary Material~\ref{sec:A_pilot_templates}). If necessary, we made minor changes to improve applicability to LLMs.

\subsubsection{Procedure} \hfill \break
\textit{Expert evaluation.} \quad Each expert rated all items from the five candidate inventories on a 4-point Likert scale (1 = not at all appropriate; 4 = fully appropriate) across three dimensions: (1) suitability to chatbot respondents, (2) suitability to the target Big Five facet, and (3) clarity of wording.

\vspace{4pt}

\noindent \textit{Pilot study.} \quad We administered the \textit{k} inventories that previously showed sufficient content validity under all seven prompt templates to each model, collecting three identitcal repetitions. This yielded a fully crossed design of 9 models $\times$ \textit{k} inventories $\times$ 7 templates $\times$ 3 repetitions.

\subsubsection{Statistical Analyses} \hfill \break
\textit{Expert study.} \quad Content validity was quantified via means of expert ratings and the scale-level content validity index \textit{S-CVI} (average proportion of experts rating $\geq 3$ across items; $>$ .80 acceptable) \cite{polit2007CVI}. Inter-rater agreement was assessed using Gwet's AC2  \cite{gwet2014interrater} (.41–.60 = moderate; .61–.80 = substantial; $\geq .81$ = almost perfect) \cite{landis1977observeragreement}, because it is more robust to low variance and high agreement among raters. The inventories with \textit{S-CVI} $\geq .80$ and at least moderate inter-rater agreement proceeded to the pilot study.

\vspace{4pt}

\noindent \textit{Pilot study.} \quad 
We selected: (1) the inventory that \textit{minimized} refusal rate, variance across repetitions, and keying sensitivity (mean score difference between positively and negatively keyed items), and \textit{maximized} correlation between the \textit{k} inventories and internal consistency (Cronbach's $\alpha$); and (2) the prompt template that \textit{minimized} refusal rate and variance across repetitions, and \textit{maximized} average correlation between templates and internal consistency.

\subsection{Phase 2: Large-Scale Administration}
To address RQ2 and RQ3, we administered the inventory and prompt template selected in Phase 1 to a large, diverse sample of LLMs. 

\subsubsection{Models} \hfill \break
We collected data from $N = 264$ LLMs, creating a representative sample that spans diverse model families, sizes, capabilities, geographic origins, licenses, and release years. Please refer to Section \ref{sec:results_phase_2_large_scale} and the GitHub repository for an overview of all models. This sample size satisfies the recommended 5:1 subject-to-item ratio for factor analysis (44 items $\rightarrow$ $N$ $\geq$ 220) \cite{gorsuch1983factor}. 

\subsubsection{Procedure} \hfill \break
We administered the winning inventory (see Supplementary Material \ref{sec:A_bfi_llm} for all items) selected in Phase 1 to all $N$ models. Models were instructed to output only a number between 1 and 5 (or 5 and 1 for reversed option order), indicating the extent to which they agree with an item. 
Via standardized API calls, we collected five identical repetitions per prompt in random order and at default temperature (see Supplementary Material \ref{sec:A_implementation} for implementation details). This produced a fully crossed design of $N$ models $\times$ 44 items $\times$ 2 option orders $\times$ 5 repetitions. Out-of-format responses and models with less than 200 valid outputs were excluded from all subsequent analyses.

\subsubsection{Statistical Analyses} \hfill

\noindent \textit{Norms}. \quad 
We tested the normality of score distributions using the Kolmogorov--Smirnov test. For each Big Five facet, we computed means, standard deviations, and percentile ranks across all $N$ models. We compared distributions with human BFI norms \citep{mcconochie2007bfi}.

\vspace{4pt}

\noindent \textit{Internal consistency}. \quad 
We computed Cronbach's $\alpha$ and McDonald's $\omega$ for each of the Big Five facets ($\geq$~.70 acceptable, $\geq$~.80 good, $\geq$~.90 excellent) \cite{george2003spss}.

\vspace{4pt}

\noindent \textit{Construct validity}. \quad
We specified a five-factor confirmatory factor analysis (CFA) with maximum-likelihood estimation to test the fit of the Big Five model. Model fit was evaluated via \textit{CFI}, \textit{TLI}, \textit{RMSEA}, and \textit{SRMR} (good fit: \textit{CFI} $\geq .95$; \textit{TLI}~$\geq 0.97$, \textit{RMSEA}, \textit{SRMR}~$\leq .05$) \cite{schumacker2004beginners}. If CFA model fit was poor, we additionally conducted exploratory factor analysis (EFA) and parallel analysis to find the optimal latent factor structure.

\subsubsection{Secondary Analyses} \hfill

\noindent \textit{Linear mixed model.} \quad
To assess which model characteristics predict personality scores, we fitted a crossed-effects linear mixed-effects model across all models, items, and repetitions. The outcome $y_{ijr}$ was the score of model $i$ on item $j$ and run $r$:

\begin{equation}
\begin{aligned}
y_{ijr} =\;& \beta_0 
+ \beta_1\,\textit{SizeGroup}_i
+ \beta_2\,\textit{ReleaseDate}_i \\
&+ \beta_3\,\textit{Reasoning}_i
+ \beta_4\,\textit{License}_i \\
&+ u_i + v_j + w_{ij} + \varepsilon_{ijr}
\end{aligned}
\end{equation}

where $u_i$, $v_j$, and $w_{ij}$ are random intercepts for models, items, and their interaction, respectively. The residual error ($\varepsilon_{ijr}$) included stochastic variation across repeated runs.
$\textit{SizeGroup}$ was a categorical predictor of parameter size (small, medium, large, or undisclosed). $\textit{ReleaseDate}$ was a continuous variable measuring time in days since model release. $\textit{Reasoning}$ and $\textit{License}$ were binary indicators of reasoning capability (yes/no) and license (open-weight/proprietary), respectively.
We fitted separate models for each individual Big Five facet.

\vspace{4pt}

\noindent \textit{Base vs.\ instruction-tuned models.} \quad
To test whether observed Big Five scores reflect alignment training rather than personality-like behavior, we compared base models with their instruction-tuned counterparts across 20 model pairs including Qwen, Gemma, Llama, Mistral, Ministral, Olmo, Granite, and Falcon models with 1B to 35B parameters. For each pair we calculate score differences ($\Delta M = \mathrm{IT} - \mathrm{base}$).
Architecture and pretraining were being held constant and only the alignment fine-tuning changed.

\vspace{4pt}

\noindent \textit{Exploratory profile comparisons.} \quad
Building up upon the linear mixed model, we compared Big Five score distributions across six model characteristics: license type (proprietary vs.\ open-weight), reasoning capability (true vs.\ false), geographic origin (US, China, Europe), number of parameters (only open-weight models), and release year. For categorical groupings we used Kruskal--Wallis tests with pairwise Mann--Whitney $U$ follow-ups.

All analyses were conducted in R (version 4.5.2) \citep{r2021} and Python (version 3.14.2) \citep{python2024, vanrossum1995python}. All materials, analyses, measures, and data are available on GitHub.


\section{Results}
\label{sec:results}

\subsection{RQ1: Do Big Five Inventories Meaningfully Describe LLMs?}
To assess whether items from human personality inventories can be used to describe LLMs, we conducted an expert evaluation and pilot study.

\subsubsection{Expert Evaluation} \hfill \break
All inventories showed high suitability to the targeted OCEAN facet ($M$ $\geq$ 3.82, \textit{S-CVI} $\geq$ .96, \textit{AC2} $\geq$ .88) and high clarity ($M$ $\geq$ 3.72, \textit{S-CVI} $\geq$ .99, \textit{AC2} $\geq$ .63), indicating that the items captured the intended personality dimensions and were clearly formulated (Tables~\ref{tab:facet_suitability} and~\ref{tab:clarity}, Supplementary Material \ref{sec:A_content_val}).
However, the suitability of items to LLM respondents differed strongly across inventories.

Table~\ref{tab:chatbot_suitability} summarizes the expert ratings of suitability to LLMs. The two established human inventories scored below the acceptable thresholds. Neither IPIP-NEO-120 (\textit{S-CVI}~=~.29) nor BFI-44 (\textit{S-CVI}~=~.54) met the required \textit{S-CVI}~$\geq$~.80.
Out of the LLM-adapted inventories, the MPI did not reach sufficient content validity (\textit{S-CVI}~=~.29). The BFI-LLM (\textit{S-CVI}~=~.93) and LMLPA (\textit{S-CVI}~=~.98) both reached high suitability ratings with substantial and almost-perfect inter-rater agreement of \textit{AC2}~=~.80 and .94, respectively.

To conclude RQ1, LLM-adapted Big Five personality test items can be valid descriptive summaries of LLM behavior, as judged by domain experts. Established human inventories showed insufficient content validity, but two out of three LLM-adapted inventories reached sufficient content validity expert ratings. Both the LMLPA and BFI-LLM proceeded to subsequent analyses.


\begin{table}[t]
\centering
\small
\begin{tabular}{lccc}
\toprule
Inventory & M & S-CVI & AC2 \\
\midrule
BFI-44 & 2.61 & .54 & .50 \\
IPIP-NEO-120 & 1.89 & .29 & .68 \\
MPI & 1.90 & .29 & .67 \\
BFI-LLM & 3.73 & .93 & .80 \\
LMLPA & 3.93 & .98 & .94 \\
\bottomrule
\end{tabular}
\par\vspace{1ex}
\footnotesize
\caption{Expert ratings of suitability to chatbot respondents. M = mean of expert ratings; S-CVI = scale-level content validity index (average proportion of items rated $\geq$ 3); AC2 = Gwet's AC2 inter-rater agreement coefficient. Higher values indicate greater LLM suitability.}
\label{tab:chatbot_suitability}
\end{table}

\subsubsection{Pilot Study} \hfill \break
We tested seven prompt templates and compared the two finalist inventories (BFI-LLM, LMLPA) on a mini-sample of nine LLMs to identify the optimal instrument and prompt.

The BFI prompt template yielded the best rating over all categories (\% \textit{refusal} = 0.97\%, variance across repetitions = 0.35, average correlation between two inventories = 0.79, keying sensitivity = 0.23, and Cronbach's $\alpha$ = 0.90) and was adopted as the prompt template in Phase~2 (Table \ref{tab:prompt_templates_results}, Supplementary Material \ref{sec:A_pilot_results}).
Out of the two finalist inventories, both the BFI-LLM and LMLPA scored low refusal rates: 3.25\% and 3.31\%, respectively. However, the BFI-LLM outperformed the LMLPA on rank-order stability (average correlation across templates ($r = .80$ vs.\ $.49$), and internal consistency ($\alpha = .82$ vs.\ $.65$) (Table \ref{tab:inventory_comparison_results}, Supplementary Material \ref{sec:A_pilot_results}). Thus, the BFI-LLM inventory was selected for Phase~2.

\subsection{RQ2: Do Big Five Inventories Capture Differences Between LLMs?} \label{sec:results_phase_2_large_scale}
To address RQ2, we conducted a large-scale administration of the best-performing Big Five inventory and analyzed LLM responses using measures of distributional properties and variability, including normality, means, standard deviations, quartiles, and variance decomposition.

\subsubsection{Descriptive Statistics and Refusal Rates} \hfill \break
The sample consisted of $N = 264$ LLMs. We collected 440 responses (44 items $\times$ 2 response option orders $\times$ 5 repetitions) per model, yielding a total of 116,160 responses, out of which 104,326 (89.81\%) were valid. 17 models were excluded from the analyses due to high refusal rate and missing values ($<$ 200 valid responses). This led to a final sample of $N = 247$ LLMs spanning 50 model families. Models originated from three main regions: the United States ($N = 113$; GPT, Llama, Gemma, Grok, Claude, Gemini, and more), China ($N = 97$; Qwen, GLM, DeepSeek, Seed, MiniMax, Kimi, MiMo, and more), and France ($N=25$; Mistral), as well as other or unspecified regions ($N=12$). It included both proprietary ($N=109$) and open-weight ($N=138$) models of varying sizes ($0.6$B $-$ $1020$B parameters, if disclosed), reasoning ($N=92$) and non-reasoning ($N=155$) models. Release dates spanned July 2023 to April 2026. 

\subsubsection{Normality} 
\begin{figure*}[t]
\centering
\includegraphics[width=\textwidth]{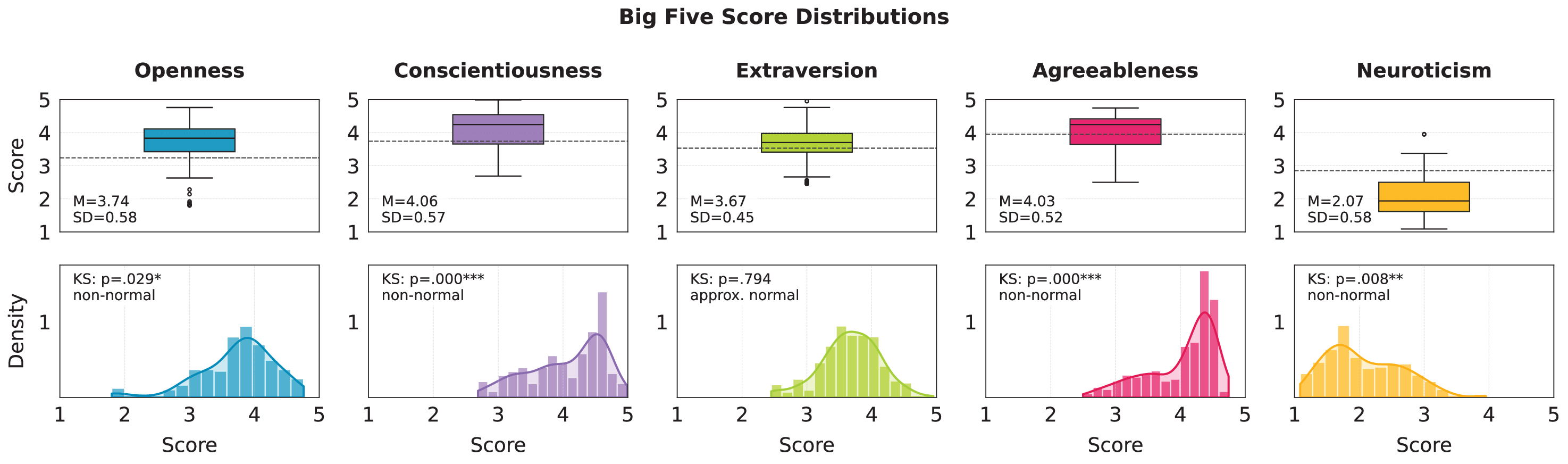}
\caption{Big Five score distributions: Boxplots and histograms across all tested models. Dotted lines are human baseline values, taken from a normative sample reported by \citet{mcconochie2007bfi} (BFI-44; $N$~=~76,413 participants; 43,540 women; age 15-65).} M~=~mean, SD~=~standard deviation, KS~=~Kolmogorov-Smirnov test.
\label{fig:rq2_ocean_distributions}
\end{figure*}

\hfill \break
Figure~\ref{fig:rq2_ocean_distributions} depicts the Big Five score distributions of all models. Kolmogorov-Smirnov tests revealed non-normality for Openness ($D$~=~0.092, $p$~$=$~.029), Conscientiousness ($D$~=~0.136, $p$~$<$~.001), Agreeableness ($D$~=~0.176, $p$~$<$~.001), and Neuroticism ($D$~=~0.105, $p$~$=$~.008). Only the facet Extraversion ($D$~=~0.041, $p$~=~.794) was approximately normally distributed. Distributions are not centered, but skewed towards the socially desirable ends of their scales.

\subsubsection{Means and Standard Deviations} \hfill \break
We compared LLM means ($M$) and standard deviations (\textit{SD}) against human normative baselines from the adult BFI-44 normative sample reported by \citet{mcconochie2007bfi} (76,413 participants; 57\% women; age 15-65). LLMs responses exhibited systematically shifted score distributions, with higher scores in Openness (LLMs: $M$~=~3.74 vs. H: $M$~=~3.24), Conscientiousness (LLMs: $M$~=~4.06 vs. H: $M$~=~3.74), Extraversion (LLMs: $M$~=~3.67 vs. H: $M$~=~3.53), and Agreeableness (LLMs: $M$~=~4.03 vs. Humans (H): $M$~=~3.95), but lower scores in Neuroticism (LLMs: $M$~=~2.07 vs. H: $M$~=~2.85). Moreover, trait variance was substantially reduced in LLMs (\textit{SD} range: 0.45--0.58 in LLMs vs. 0.65--0.89 in humans). 

\subsubsection{Variance Decomposition} \hfill \break
The random-effects variance decomposition of the mixed-effects model confirmed this pattern. Between-model differences explained the smallest proportion of the total variance in OCEAN scores (Table \ref{tab:variance_decomposition}), with only 7\% to 17\% of total score variance explained by model differences. Item-level variance and model-item interaction explained up to 33\% and 41\% of the total variance, respectively. Models responded differently to specific items but overall still exhibited similar personality profiles. Finally, residual variance accounted for up to 41\% of total variance, suggesting that there is unexplained variation across individual responses, not captured by model or item effects. Overall, response variability was primarily driven by item formulation, item-specific interpretation, and randomness across repetitions, rather than than stable differences between models.

\begin{table}[t]
\centering
\small
\begin{tabular}{ccccc}
\toprule
  & $\hat{\sigma}^2_{model}$ & $\hat{\sigma}^2_{item}$ & $\hat{\sigma}^2_{model:item}$ & $\hat{\sigma}^2_{res}$ \\
\midrule
\textbf{O} & 0.22 (17\%) & 0.17 (13\%) & 0.38 (30\%) & 0.53 (41\%) \\
\textbf{C} & 0.14 (10\%) & 0.27 (19\%) & 0.52 (37\%) & 0.47 (33\%) \\
\textbf{E} & 0.09 (7\%)  & 0.20 (14\%) & 0.57 (41\%) & 0.53 (38\%) \\
\textbf{A} & 0.12 (7\%)  & 0.52 (33\%) & 0.53 (33\%) & 0.43 (27\%) \\
\textbf{N} & 0.12 (9\%)  & 0.20 (16\%) & 0.45 (36\%) & 0.48 (38\%) \\
\bottomrule
\end{tabular}
\par\vspace{1ex}
\footnotesize
\caption{Variance decomposition of the linear mixed models for each Big Five trait. Percentages indicate the proportion of explained variance by each component compared to the total variance. O~=~Openness, C~=~Conscientiousness, E~=~Extraversion, A~=~Agreeableness, N~=~Neuroticism.}
    \label{tab:variance_decomposition}
\end{table}

The analyses to address RQ2 provide strong evidence that Big Five scores do not meaningfully differ across LLMs. Score distributions were shifted toward socially desirable ends, characterized by elevated Openness, Conscientiousness, Extraversion, and Agreeableness, and reduced Neuroticism. Between-model variance accounted for only 7\% to 17\% of the total variance, while item identity, model--item interactions, and residual stochastic variation accounted for the majority of observed differences in OCEAN scores. Compared to human norms, LLM responses also exhibited reduced variance and non-normal distributions for four of the five traits. These results indicate that Big Five inventories do not capture meaningful differences between LLMs.  

\subsection{RQ3: Do Big Five Inventories Reflect The Big Five Structure in LLMs?}
To tackle RQ3, we assessed whether LLMs’ Big Five responses reflect coherent, internal factors through measures of internal consistency and factor analysis.

\vspace{8pt}

\subsubsection{Internal Consistency} \hfill \break
Cronbach’s $\alpha$ reached good internal consistency for Openness ($\alpha~=~.86$), Conscientiousness ($\alpha~=~.81$) and Neuroticism ($\alpha~=~.81$). Agreeableness ($\alpha~=~.76$) showed acceptable internal consistency and Extraversion fell below conventional acceptable levels ($\alpha~=~.61$). McDonald’s $\omega$ estimates resulted in ultra-Heywood solutions for all scales except Conscientiousness and are therefore not reported. 

\vspace{8pt}

\subsubsection{Factor Structure} \hfill \break
We conducted a five-factor confirmatory factor analysis (CFA) at the item level, using a model-by-item data matrix (247 models × 44 BFI items). Each model contributed one response per item, aggregated over the two option orders and five repetitions.
The CFA demonstrated poor model fit ($\chi^2(892)$~=~6169.49, $p < .001$, \textit{CFI}~=~.53, \textit{TLI}~=~.50, \textit{RMSEA}~=~.17, \textit{SRMR}~=~.32), indicating substantial deviation from the hypothesized five-factor personality structure. Thus, the standard Big Five model does not adequately capture the covariance structure of the LLMs' responses.

\begin{figure}[t]
    \centering
    \includegraphics[width=0.95\columnwidth]{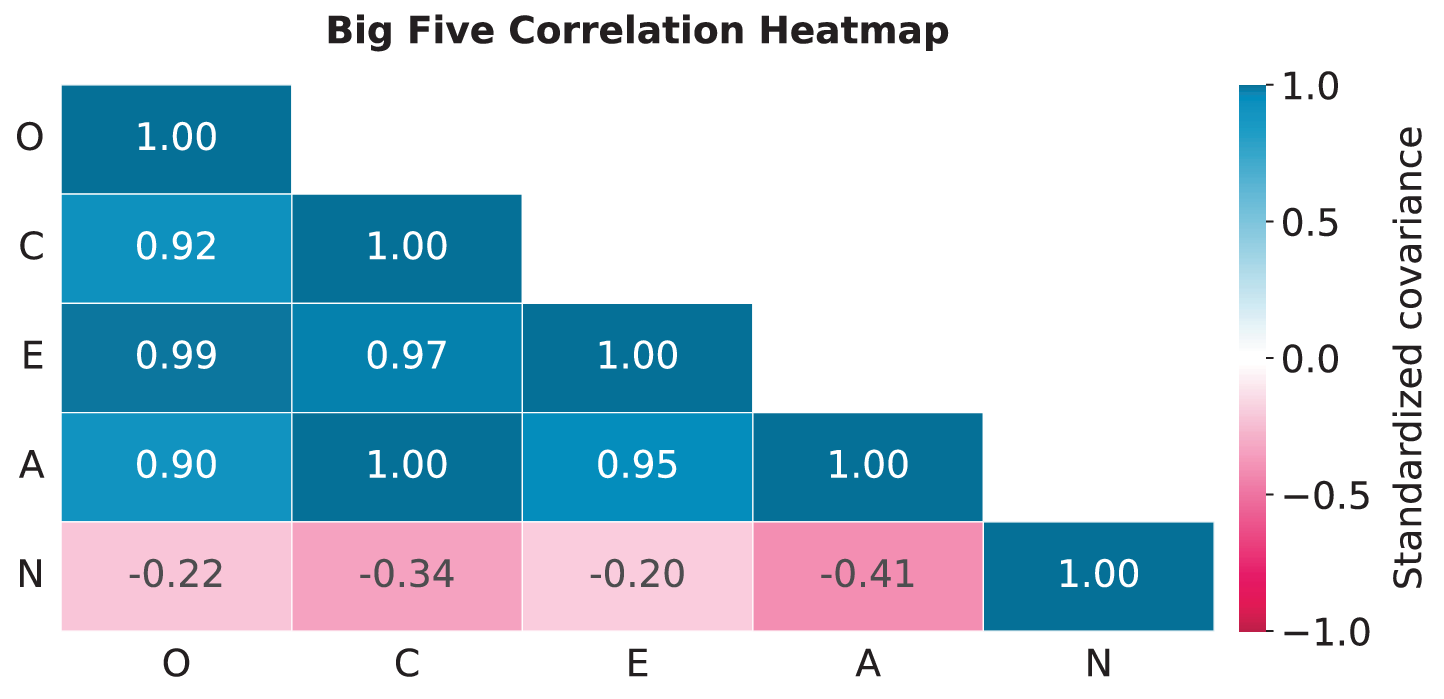}
    \caption{Correlation heatmap retrieved from CFA with a five-factor solution expected from the Big Five. O~=~Openness, C~=~Conscientiousness, E~=~Extraversion, A~=~Agreeableness, N~=~Neuroticism.}
    \label{fig:rq3_correlation}
\end{figure}

Between-facet correlations for Extraversion, Agreeableness, Conscientiousness, and Openness were very high ($r~\geq~.90$, all $p <$ .001), indicating near-complete collinearity rather than separable dimensions (Figure \ref{fig:rq3_correlation}). Neuroticism showed weak to moderate negative covariances with the remaining four traits ($r$~=~-.20 to -.41, $p$~=~.090 to .002). 

Furthermore, reverse-coded items frequently exhibited weak or negative loadings on their respective latent OCEAN factors, despite being recoded prior to CFA. On average, reverse-coded showed substantially lower factor loadings ($M$~=~0.14, \textit{SD}~=~0.27), with 6 out of 16 (37.5\%) failing to reach statistical significance. In contrast, positively-keyed items showed higher average factor loadings ($M$~=~0.76, \textit{SD}~=~0.28) with only 1 out of 28 items (3.6\%) being non-significant. These findings provide further evidence that Big Five scores are influenced more by item formulation characteristics, such as negation and reverse wording, than by model-specific latent factors.

We additionally conducted an exploratory factor analysis (EFA). The scree plot with elbow-criterion suggested a two-factor solution (Figure \ref{fig:scree_plot}, Supplementary Material \ref{sec:factor_structure}), which achieved improved but still inadequate model fit ($\chi^2(859)$ = 2789.37, $p < .001$, \textit{TLI} = .83, \textit{RMSEA} = .10).

In summary, the internal consistency analyses indicate that all facets except Extraversion achieved at least acceptable reliability, with most reaching good levels. However, the CFA results show that LLM responses do not match the human-derived five-factor structure of personality. Moreover, even the best-fitting EFA solution failed to adequately represent the structure underlying LLM responses. These findings suggest that self-reported Big Five responses from LLMs do not organize into coherent, interpretable personality dimensions analogous to those observed in humans.

\subsection{What Are Big Five Inventories Measuring?}
The results of RQ1 -- RQ3 suggest that Big Five inventories do not measure human-like personality traits in LLMs. We therefore conduct additional analyses to investigate which patterns and artifacts may explain the structure observed in LLM responses.

\subsubsection{Base vs.\ Instruction-Tuned Models} \hfill \break
To investigate whether alignment training is associated with differences in Big Five scores, we compared the responses of 20 base models with their instruction-tuned counterparts, while holding architecture and pretraining constant. Table~\ref{tab:base_instruct_diff} reports aggregated mean differences and standard deviations across the 20 pairs.
Instruction-tuned variants scored higher than their base counterparts on Openness, Conscientiousness, Extraversion, and Agreeableness, and lower on Neuroticism. This pattern suggests that instruction tuning is associated with shifts toward more socially desirable response profiles in Big Five scores.

\begin{table}[h]
\centering
\small
\begin{tabular}{lrc}
\toprule
Trait & $\Delta M$ & \textit{SD} \\
\midrule
Openness             &  0.12 & 0.75 \\
Conscientiousness    &  0.44 & 0.59 \\
Extraversion         &  0.24 & 0.64 \\
Agreeableness        &  0.57 & 0.58 \\
Neuroticism          & $-$0.47 & 0.54 \\
\bottomrule
\end{tabular}
\par\vspace{1ex}
\footnotesize
\caption{Comparison of Big Five scores between base and instruction-tuned (IT) models. $\Delta M$~=~score of the IT version minus score of the base version, averaged across $N~=~20$ model pairs. Positive values indicate higher scores in IT models. \textit{SD}~=~standard deviation across pairs.}
\label{tab:base_instruct_diff}
\end{table}

\subsubsection{Predictors of Personality Scores} 
\begin{figure*}[t]
  \centering
  \includegraphics[width=\textwidth]{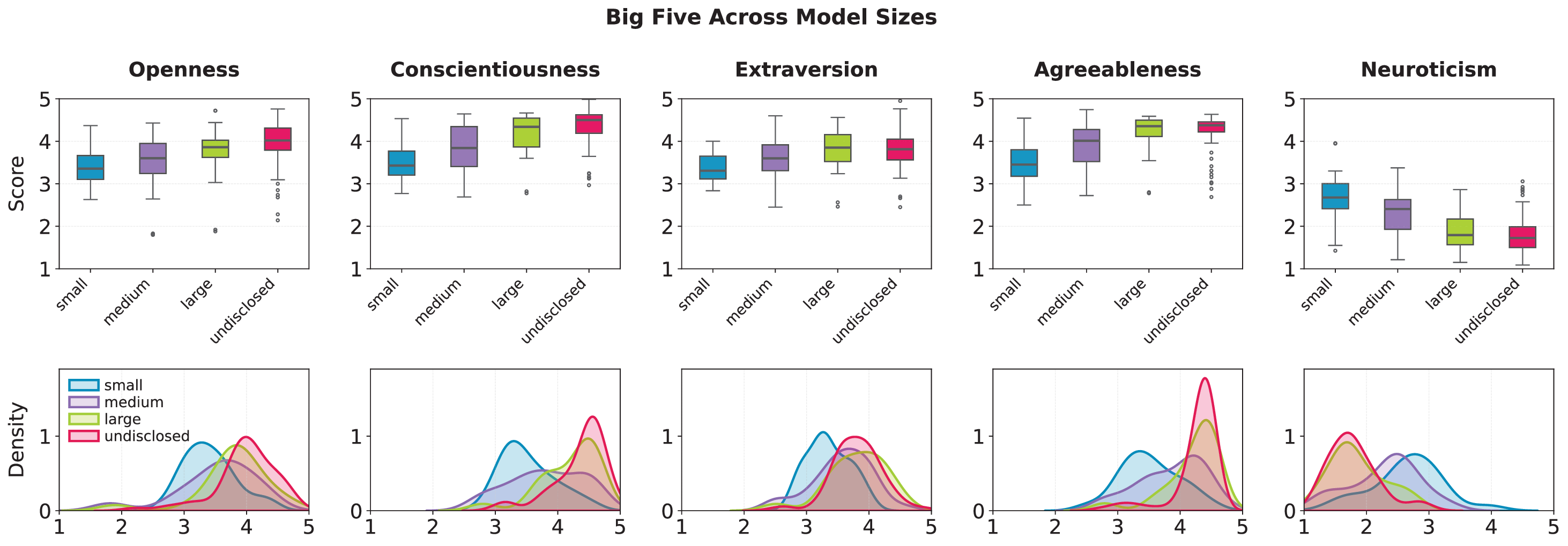}
  \caption{Big Five score distributions by model parameter scale (small ($<$10B), medium (10-100B), large ($>$100B), undisclosed).}
  \label{fig:sizes}
  \end{figure*}

\hfill \break
Table~\ref{tab:lmer_ocean} in Supplementary Material \ref{sec:A_subgroup} reports the fixed-effect estimates of all linear mixed models. Only model size significantly predicted all traits: small models scored significantly lower than large and undisclosed models on Openness, Conscientiousness, Extraversion, and Agreeableness, but higher on Neuroticism (Figure \ref{fig:sizes}). Medium models exhibited a similar but weaker trend for all traits except Openness.

Reasoning capability predicted higher Conscientiousness and Agreeableness, and lower Neuroticism, but did not significantly predict Extraversion or Openness (Figure \ref{fig:reasoning}, Supplementary Material \ref{sec:A_subgroup}). When controlling for all other predictors, release date negatively predicted Openness, Conscientiousness, and Agreeableness, and predicted higher Neuroticism. 
Although Figure~\ref{fig:release_date_regression} (Supplementary Material \ref{sec:A_subgroup}) shows a reverse trend toward more socially desirable scores in more recently released models, this pattern could be largely driven by the growing size, instruction-tuning, and reasoning capabilities of models in more recent releases shown in Figure~\ref{fig:size_regression} (Supplementary Material \ref{sec:A_subgroup}). License was non-significant across all traits.

\subsubsection{Geographic Origin} \hfill \break
Figure~\ref{fig:countries} in Supplementary Material \ref{sec:A_subgroup} depicts the Big Five score distributions for models grouped by geographic origin: US, China, and France (represented by Mistral only). Kruskal--Wallis tests revealed significant differences in Agreeableness ($H = 22.10$, $p < .001$) and Conscientiousness ($H = 15.47$, $p = .019$): both Chinese and US models scored higher than French models. Chinese models additionally scored higher than US models on Agreeableness.
Neuroticism showed the largest difference ($H = 34.46$, $p < .001$): French models scored higher than US models, which scored higher than Chinese models. Extraversion ($H = 3.38$, $p = .185$) and Openness ($H = 2.11$, $p = .348$) showed no significant differences across regions.

\subsubsection{Model Family} 
\begin{figure}[h]
    \centering
    \includegraphics[width=0.95\columnwidth]{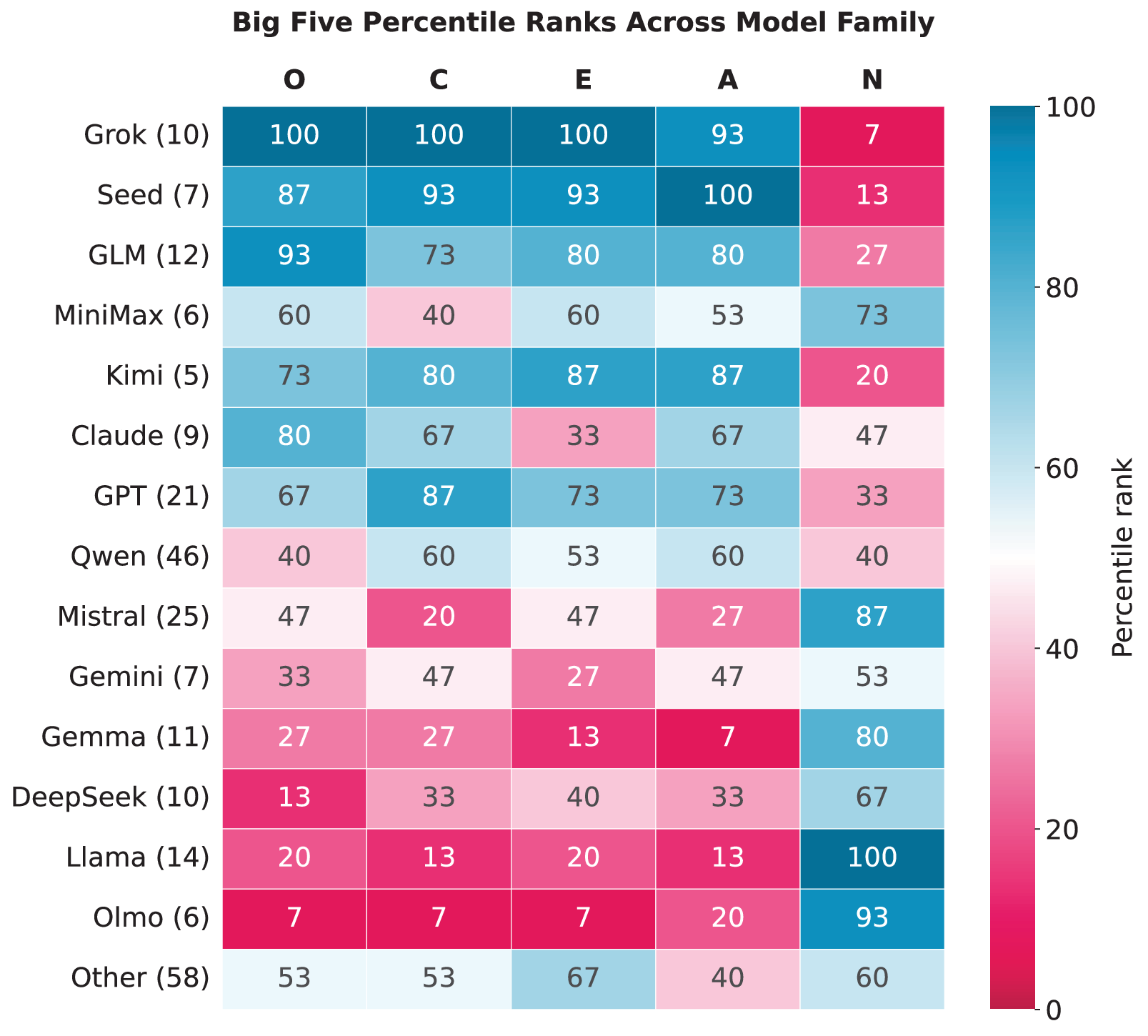}
    \caption{Percentile ranks (percentage of model families with lower scores) of Big Five (OCEAN) trait scores across LLM families. A percentile rank of 50 indicates the median; blue indicates values above; red indicates values below. (.) denotes the number of models averaged for each model family (minimum five).  O~=~Openness, C~=~Conscientiousness, E~=~Extraversion, A~=~Agreeableness, N~=~Neuroticism.}
    \label{fig:families_heatmap}
\end{figure}

\hfill \break
We found differences in Big Five profiles across model families (Figure~\ref{fig:families_heatmap}). Out of all models, Grok models scored highest on Openness, Conscientiousness, and Extraversion (Percentile rank (\textit{PR}) = 100, respectively), and lowest on Neuroticism (\textit{PR}~=~7).
Seed models scored highest on Agreeableness (\textit{PR}~=~100). Olmo models scored lowest on Openness, Conscientiousness, and Extraversion (\textit{PR}~=~7, respectively). Gemma scored lowest in Agreeableness (\textit{PR}~=~7) and Llama scored highest on Neuroticism (\textit{PR}~=~100).

Despite these differences, between-model variability accounted for only 7\% to 17\% of total variance. This means that even though there are some differences between model families, these differences are small relative to other sources of variance, such as item formulation. As the Big Five factor structure was not confirmed by CFA, these trait labels and achieved scores should be interpreted with caution. 


\section{Discussion}
\label{sec:discussion}

This paper examined whether three foundational characteristics of personality traits hold when applying Big Five inventories to LLMs: whether they (a) appropriately describe LLMs, (b) capture meaningful differences between models, and (c) reflect latent factors consistent with human personality. Across all three dimensions, our findings provide limited support for applying human personality frameworks to LLMs. We discuss the implications for AI governance and the challenge of characterizing LLM behavior.

\paragraph{Content Validity Cannot Be Assumed} \hfill \break
The two widely used human inventories (BFI-44 and IPIP-NEO) showed insufficient content validity when applied to LLMs. Two LLM-adapted inventories achieved acceptable levels. Interestingly, the MPI, despite being designed specifically for LLMs, performed similarly to human inventories. The MPI is based on IPIP items, but its adaptation process to LLMs is not described \cite{jiang2023evaluating}.

These findings suggest that previous studies applying human Big Five inventories to LLMs without establishing content validity may have introduced measurement error. Before interpreting model scores as personality traits, measurement instruments must first be shown to adequately represent the target population. Consequently, governance claims that characterize models as “agreeable” or “conscientious” based on such scores should be interpreted cautiously.

\paragraph{Big Five Scores Capture Few Model Differences} \hfill \break
LLMs exhibited substantially lower variability in Big Five scores than human normative samples. This finding aligns with prior work showing reduced variance in personality scores across LLMs \cite{huang2024revisiting}. In line with \citeauthor{salecha2024socialdesirability} (\citeyear{salecha2024socialdesirability}), models generally showed similarly high scores on socially desirable dimensions, with only small differences across models and factors. The observed variance patterns indicate that Big Five scores capture relatively little variation attributable to model differences. Instead, response patterns appear to be strongly associated with factors such as item formulation and sampling variability. Thus, differences in Big Five scores should \textit{not} automatically be interpreted as differences in underlying model characteristics.

\paragraph{Factor Structure Fails to Replicate Human Personality} \hfill \break
Internal consistency was generally acceptable, suggesting that items from the same facet measure a similar underlying construct. However, CFA did not recover the five-factor structure observed in humans. Most OCEAN dimensions were highly correlated ($r \geq .90$), suggesting limited separation between the constructs. EFA produced a two-factor solution, but model fit remained poor. Overall, LLM responses to Big Five inventories do not appear to organize according to the same latent structure as human personality.

This finding is consistent with prior studies using different inventories and sampling procedures. 
\citeauthor{peereboom2025cognitive} (\citeyear{peereboom2025cognitive}) found no meaningful latent structure in LLM responses using HEXACO with temperature sampling, while \citeauthor{suehr2024challenging} (\citeyear{suehr2024challenging}) did not recover the expected five-factor structure using BFI-2 with persona prompting.

Reverse-coded items provide additional insight into these results. Despite recoding prior to analysis, negatively keyed items showed substantially weaker or negative loadings on their intended factors. This pattern suggests that LLMs might systematically process positively and negatively keyed items differently. It may also contribute to why Neuroticism stands out as a relative distinct factor, as it is largely defined by negatively worded items. These observations are consistent with known challenges in LLM processing negations \cite{vrabcova2025negation} and prior findings that reverse-coded personality items loaded on a separate factors than positively coded items \cite{salecha2024socialdesirability}, or LLMs responding affirmatively to both positively and negatively keyed items \cite{suehr2024challenging}.

\paragraph{Instruction-Tuning and Socially Desirable Patterns} \hfill \break
Instruction-tuned models consistently exhibit higher scores on Openness, Conscientiousness, Agreeableness, and Extraversion, and lower scores on Neuroticism compared to their base counterparts. These differences are most pronounced for Agreeableness and Conscientiousness, and consistent with a shift toward more socially desirable response patterns after instruction-tuning. These patterns are consistent across model families and might possibly explain why recent and bigger models also score higher on these social desirable traits, as release date and size might be correlated with more instruction-tuned models.
As instruction-tuned models are designed to be helpful, honest, and harmless, Big Five scores may capture alignment of models to being a helpful assistant rather than model-specific characteristics.

\paragraph{Differences in Region and Family Raise Governance Questions} \hfill \break
We observed systematic differences in Big Five scores across model families and geographic origins. For example, French models scored relatively higher on Neuroticism and lower on Agreeableness and Conscientiousness compared to US- and China-based model families. Grok models showed among the highest scores on socially desirable dimensions. However, these differences should not be interpreted as evidence of distinct model personalities. They may reflect variation in training data, alignment practices, model size, release strategy, or safety objectives. Similarly, models such as Olmo, Mistral, and Llama, which are often more open-weight and less heavily instruction-tuned, showed weaker socially desirable profiles. These observations are consistent with prior work reporting stronger social desirability patterns in larger and more instruction-tuned models \cite{tosato2025persistent}. Overall, personality-like response patterns may reflect institutional and technical design choices rather than intrinsic properties of the models.

This distinction has important governance implications. If Big Five scores partly capture differences in alignment practices or cultural and institutional design choices, cross-model comparisons based on these scores may be difficult to interpret. Governance frameworks should therefore treat such measurements as indicators of training and interaction patterns rather than as direct measures of model personality.

\paragraph{Rethinking the Use of Human Personality Frameworks For LLMs} \hfill \break
Our findings suggest that Big Five inventories should not be interpreted as measuring human-like personality traits in LLMs. Applying human personality categories to artificial systems risks attributing psychological properties that have not been empirically validated. Such anthropomorphic interpretations may contribute to users perceiving models as having intentions, emotions, or understanding, with potential implications for persuasion and manipulation \cite{nass1993anthropomorphism,peter2025benefits}.

Rather than asking whether LLMs have human personality traits, evaluations should focus on identifying characteristics that are meaningful to describe LLMs. Psychological frameworks provide useful evaluation protocols and psychometric methods, but they require validation before being applied beyond the populations for which they were developed.

\paragraph{Toward Valid Behavioral Characterization of LLMs} \hfill \break
These findings highlight the need for evaluation frameworks that are developed for LLMs, rather than adopting human constructs without validation. Future work should identify LLM-native constructs in terms of observable model behavior rather than exclusively relying on self-reports of LLMs.

We hypothesize that LLMs might have inherent characteristics that satisfy the six defining criteria of personality - even though they are not the Big Five. Potential candidates could include characteristics such as sycophancy, prompt sensitivity, instruction-following consistency, robustness to context changes, and resistance to manipulation. These constructs could satisfy some of the defining criteria of personality-like characteristics while avoiding unsupported claims about internal states. Importantly, they may map directly onto practical governance concerns, including safety calibration, reliability, and deployment robustness.

Developing and validating LLM-specific frameworks using systematic evaluation pipelines (see Section~\ref{sec:eval_pipeline}), should therefore be a priority. As AI governance increasingly relies on technical evaluations, including assessments under the EU AI Act, measurement approaches require transparent methodology and empirical validation. Human personality frameworks that lack evidence of construct validity for LLMs are unlikely to provide such assessments.

\paragraph{Limitations} \hfill \break
Our work is not exempt of limitations. Content validity was assessed by three expert raters, the established minimum ~\citep{lynn1986determination} but a small panel nonetheless. Two of the three raters were authors of this study, which may introduce potential bias in the reported agreement estimates.

The normative sample covers models available through public APIs as of early 2026 and may not generalize to future architectures or modalities, such as vision-language models. Geographic comparisons are limited by the small number of European models, and results for this region should be treated as illustrative. The comparison between instruction-tuned and base models might partly reflect prompt sensitivity rather than the effect of instruction tuning alone, as base models were mostly local models prompted with an additional JSON constraint. 

All inventory items, prompt templates, and model administrations were conducted in English. Findings may not generalize to other languages. Our analyses rely on self-reports, a format that may not capture actual behavioral tendencies. 
The independence (iid) assumption underlying our dataset is imperfect, models within the same family share architecture and training data, meaning responses may be correlated. 

Our evaluation focuses on the first three characteristics of the proposed evaluation pipeline. While these might provide important evidence against applying human Big Five inventories to LLMs, future work should examine temporal stability, cross-situational consistency, and behavioral relevance through longitudinal assessments and behavioral validation.


\section{Conclusion}
This paper examined whether Big Five inventories can be validly applied to LLMs by testing the assumptions underlying this transfer. While LLM-adapted items achieved sufficient content validity, Big Five scores showed few differences between models and failed to reproduce the human five-factor structure. 
Instruction-tuned model variants also exhibited more socially desirable profiles than their base counterparts. These findings suggest that Big Five inventories do not capture human-like personality traits in LLMs. 

Because personality scores are increasingly used to benchmark, compare, and govern AI systems, relying on unvalidated human constructs risks misleading interpretations. Future evaluation frameworks should prioritize LLM-native constructs, validated measurement instruments, and behavioral assessments rather than human self-report analogues.


\section*{Acknowledgments}
This work has been supported by a nominal grant received at the ELLIS Unit Alicante Foundation from the Regional Government of Valencia in Spain (Resolución de la Conselleria de Industria, Turismo, Innovación y Comercio, Dirección General de Innovación) and by Intel Corporation. K.Z. was supported by the Bank Sabadell Foundation. We thank Ágnes Magyary for her help with the content validity study, and we thank Thilo Hagendorff for giving feedback on the paper.


\bibliography{library}

@article{costa1986personality,
title = {Personality stability and its implications for clinical psychology},
journal = {Clinical Psychology Review},
volume = {6},
number = {5},
pages = {407-423},
year = {1986},
note = {Special Issue Personality Assessment in the 80's: Issues and Advances},
issn = {0272-7358},
doi = {https://doi.org/10.1016/0272-7358(86)90029-2},
url = {https://www.sciencedirect.com/science/article/pii/0272735886900292},
author = {Paul T. Costa and Robert R. McCrae},
}

@article{zheng2025LMLPA,
author = {Zheng, Jingyao and Wang, Xian and Hosio, Simo and Xu, Xiaoxian and Lee, Lik-Hang},
title = {LMLPA: Language Model Linguistic Personality Assessment},
journal = {Computational Linguistics},
volume = {51},
number = {2},
pages = {599-640},
year = {2025},
month = {06},
issn = {0891-2017},
doi = {10.1162/coli_a_00550},
url = {https://doi.org/10.1162/coli_a_00550}
}

@inproceedings{jiang2023evaluating,
author = {Jiang, Guangyuan and Xu, Manjie and Zhu, Song-Chun and Han, Wenjuan and Zhang, Chi and Zhu, Yixin},
title = {Evaluating and inducing personality in pre-trained language models},
year = {2023},
publisher = {Curran Associates Inc.},
address = {Red Hook, NY, USA},
booktitle = {Proceedings of the 37th International Conference on Neural Information Processing Systems},
articleno = {466},
numpages = {22},
location = {New Orleans, LA, USA},
series = {NIPS '23}
}

@article{salecha2024socialdesirability,
author = {Salecha, Aadesh and Ireland, Molly E and Subrahmanya, Shashanka and Sedoc, João and Ungar, Lyle H and Eichstaedt, Johannes C},
title = {Large language models display human-like social desirability biases in Big Five personality surveys},
journal = {PNAS Nexus},
volume = {3},
number = {12},
pages = {pgae533},
year = {2024},
month = {12},
issn = {2752-6542},
doi = {10.1093/pnasnexus/pgae533},
url = {https://doi.org/10.1093/pnasnexus/pgae533}
}

@article{huang2025designing,
author = {Muhua Huang and Xijuan Zhang and Christopher Soto and James Evans},
title ={Designing AI-Agents With Personalities: A Psychometric Approach},
journal = {Personality Science},
volume = {7},
number = {},
pages = {27000710251406471},
year = {2026},
doi = {10.1177/27000710251406471},
URL = {https://doi.org/10.1177/27000710251406471},
eprint = {https://doi.org/10.1177/27000710251406471},
}

@misc{karra2023estimating,
title = {Estimating the Personality of White-Box Language Models},
author = {Saketh Reddy Karra and Son The Nguyen and Theja Tulabandhula},
year = {2023},
eprint = {2204.12000},
archivePrefix = {arXiv}
}

@misc{bai2025scaling,
title = {Scaling Law in LLM Simulated Personality: More Detailed and Realistic Persona Profile Is All You Need},
author = {Yuqi Bai and Tianyu Huang and Kun Sun and Yuting Chen},
year = {2025},
eprint = {2510.11734},
archivePrefix = {arXiv}
}

@inproceedings{lee2025llmsdistinct,
title = "Do {LLM}s Have Distinct and Consistent Personality? {TRAIT}: Personality Testset designed for {LLM}s with Psychometrics",
author = "Lee, Seungbeen  and
  Lim, Seungwon  and
  Han, Seungju  and
  Oh, Giyeong  and
  Chae, Hyungjoo  and
  Chung, Jiwan  and
  Kim, Minju  and
  Kwak, Beong-woo  and
  Lee, Yeonsoo  and
  Lee, Dongha  and
  Yeo, Jinyoung  and
  Yu, Youngjae",
booktitle = "Findings of the {A}ssociation for {C}omputational {L}inguistics: {NAACL} 2025",
month = apr,
year = "2025",
address = "Albuquerque, New Mexico",
publisher = "Association for Computational Linguistics",
url = "https://aclanthology.org/2025.findings-naacl.469/",
doi = "10.18653/v1/2025.findings-naacl.469",
pages = "8412--8452",
ISBN = "979-8-89176-195-7",
}

@inproceedings{li2025evalpsychometrics,
title = {Evaluating Large Language Models with Psychometrics},
author = {Yuan Li and Yue Huang and Hongyi Wang and Ying Cheng and Xiangliang Zhang and James Zou and Lichao Sun},
booktitle = {Large Language Models for Scientific and Societal Advances},
year = {2025},
url = {https://openreview.net/forum?id = OSsQ5AUz6X}
}

@inproceedings{suehr2024challenging,
author = {S\"{u}hr, Tom and Dorner, Florian E. and Samadi, Samira and Kelava, Augustin},
title = {Challenging the Validity of Personality Tests for Large Language Models},
year = {2025},
isbn = {9798400721403},
publisher = {Association for Computing Machinery},
address = {New York, NY, USA},
url = {https://doi.org/10.1145/3757887.3763016},
doi = {10.1145/3757887.3763016},
booktitle = {Proceedings of the 5th ACM Conference on Equity and Access in Algorithms, Mechanisms, and Optimization},
pages = {74–81},
numpages = {8},
location = {
},
series = {EAAMO '25}
}

@misc{romero2023llmssplit,
title = {Do GPT Language Models Suffer From Split Personality Disorder? The Advent Of Substrate-Free Psychometrics},
author = {Romero, Peter and Fitz, Stephen and Nakatsuma, Teruo},
year = {2024},
eprint = {2408.07377},
archivePrefix = {arXiv},
}

@misc{tosato2025persistent,
title = {Persistent Instability in LLM's Personality Measurements: Effects of Scale, Reasoning, and Conversation History},
author = {Tommaso Tosato and Saskia Helbling and Yorguin-Jose Mantilla-Ramos and Mahmood Hegazy and Alberto Tosato and David John Lemay and Irina Rish and Guillaume Dumas},
year = {2025},
eprint = {2508.04826},
archivePrefix = {arXiv}
}

@article{mccrae1992introduction,
title = {An {Introduction} to the {Five}‐{Factor} {Model} and {Its} {Applications}},
volume = {60},
copyright = {http://onlinelibrary.wiley.com/termsAndConditions\\#vor},
issn = {0022-3506, 1467-6494},
url = {https://onlinelibrary.wiley.com/doi/10.1111/j.1467-6494.1992.tb00970.x},
doi = {10.1111/j.1467-6494.1992.tb00970.x},
language = {en},
number = {2},
urldate = {2026-01-26},
journal = {Journal of Personality},
author = {McCrae, Robert R. and John, Oliver P.},
month = jun,
year = {1992},
pages = {175--215}
}

@book{larsen2017personalitypsychology,
title = {Personality Psychology: Domains of Knowledge about Human Nature},
author = {Larsen, Randy J. and Buss, David M. and King, David B. and Ensley, Carolyn E.},
year   = {2017},
publisher = {McGraw-Hill Education},
address = {Toronto, Ontario},
note = {Canadian edition; based on \emph{Personality Psychology}, 5th ed. (2014)}
}

@book{matthews2003personalitytraits,
title  = {Personality Traits},
author = {Matthews, Gerald and Deary, Ian J. and Whiteman, Martha C.},
edition = {2},
year = {2003},
publisher = {Cambridge University Press},
address = {Cambridge, United Kingdom},
isbn = {0521831075},
}

@misc{huang2024revisiting,
title = {Revisiting the Reliability of Psychological Scales on Large Language Models},
author = {Jen{-}tse Huang and Wenxiang Jiao and Man Ho Lam and Eric John Li and Wenxuan Wang and Michael R. Lyu},
year = {2024},
eprint = {2305.19926},
archivePrefix = {arXiv}
}

@article{bodroza2024personality,
author = {Bodroža, Bojana  and Dinić, Bojana M.  and Bojić, Ljubiša },
title = {Personality testing of large language models: limited temporal stability, but highlighted prosociality},
journal = {Royal Society Open Science},
volume = {11},
number = {10},
pages = {240180},
year = {2024},
doi = {10.1098/rsos.240180},
URL = {https://royalsocietypublishing.org/doi/abs/10.1098/rsos.240180}
}

@misc{sorokovikova2024llmssimulate,
title = {LLMs Simulate Big Five Personality Traits: Further Evidence}, 
author = {Aleksandra Sorokovikova and Natalia Fedorova and Sharwin Rezagholi and Ivan P. Yamshchikov},
year = {2024},
eprint = {2402.01765},
archivePrefix = {arXiv}
}

@misc{hilliard2024eliciting,
title = {Eliciting Personality Traits in Large Language Models},
author = {Airlie Hilliard and Cristian Munoz and Zekun Wu and Adriano Soares Koshiyama},
year = {2024},
eprint = {2402.08341},
archivePrefix = {arXiv},
}

@inproceedings{hartley2025personality,
title = {How Personality Traits Shape {LLM} Risk-Taking Behaviour},
author = {Hartley, John  and
Hamill, Conor Brian  and
Seddon, Dale  and
Batra, Devesh  and
Okhrati, Ramin  and
Khraishi, Raad},
booktitle = {Findings of the Association for Computational Linguistics: ACL 2025},
month = jul,
year = {2025},
address = {Vienna, Austria},
publisher = {Association for Computational Linguistics},
url = {https://aclanthology.org/2025.findings-acl.1085/},
doi = {10.18653/v1/2025.findings-acl.1085},
pages = {21068--21092},
ISBN = {979-8-89176-256-5},
}

@article{pellert2024AIpsychometrics,
author = {Max Pellert and Clemens M. Lechner and Claudia Wagner and Beatrice Rammstedt and Markus Strohmaier},
title  = {AI Psychometrics: Assessing the Psychological Profiles of Large Language Models Through Psychometric Inventories},
journal = {Perspectives on Psychological Science},
volume = {19},
number = {5},
pages = {808-826},
year = {2024},
doi = {10.1177/17456916231214460},
note  = {PMID: 38165766},
URL = {https://doi.org/10.1177/17456916231214460}
}

@article{xie2025aipsychobench,
author = {Xie, Wei and Wang, Zhenhua and Ma, Shuoyoucheng and Sun, Xiaobing and Chen, Kai and Wang, Enze and Liu, Wei and Tong, Hanying},
title = {AIPsychoBench: Understanding the Psychometric Differences Between LLMs and Humans},
journal = {Topics in Cognitive Science},
volume = {18},
number = {2},
pages = {e70041},
doi = {https://doi.org/10.1111/tops.70041},
url = {https://onlinelibrary.wiley.com/doi/abs/10.1111/tops.70041},
eprint = {https://onlinelibrary.wiley.com/doi/pdf/10.1111/tops.70041},
year = {2026}
}

@inproceedings{shu2024dontneed,
title = {You don{'}t need a personality test to know these models are unreliable: Assessing the Reliability of Large Language Models on Psychometric Instruments},
author = {Shu, Bangzhao  and
Zhang, Lechen  and
Choi, Minje  and
Dunagan, Lavinia  and
Logeswaran, Lajanugen  and
Lee, Moontae  and
Card, Dallas  and
Jurgens, David},
booktitle = {Proceedings of the 2024 Conference of the North American Chapter of the Association for Computational Linguistics: Human Language Technologies (Volume 1: Long Papers)},
month = jun,
year = {2024},
address = {Mexico City, Mexico},
publisher = {Association for Computational Linguistics},
url = {https://aclanthology.org/2024.naacl-long.295/},
doi = {10.18653/v1/2024.naacl-long.295},
pages = {5263--5281},
}

@article{goldberg1992bigfive,
author = {Goldberg, Lewis R.},
title  = {The development of markers for the Big-Five factor structure},
journal = {Psychological Assessment},
year = {1992},
volume = {4},
number = {1},
pages  = {26--42},
doi = {10.1037/1040-3590.4.1.26},
publisher = {American Psychological Association}
}

@misc{serapiogarcia2025personality,
title = {Personality Traits in Large Language Models},
author = {Greg Serapio-García and Mustafa Safdari and Clément Crepy and Luning Sun and Stephen Fitz and Peter Romero and Marwa Abdulhai and Aleksandra Faust and Maja Matarić},
year = {2025},
eprint = {2307.00184},
archivePrefix = {arXiv}
}

@inproceedings{frisch2024LLMagents,
title = {"LLM" Agents in Interaction: Measuring Personality Consistency and Linguistic Alignment in Interacting Populations of Large Language Models},
author = {Frisch, Ivar  and
Giulianelli, Mario},
booktitle = {Proceedings of the 1st Workshop on Personalization of Generative AI Systems (PERSONALIZE 2024)},
month = mar,
year = {2024},
address = {St. Julians, Malta},
publisher = {Association for Computational Linguistics},
url = {https://aclanthology.org/2024.personalize-1.9/},
pages = {102--111},
}

@misc{han2025personalityillusion,
title = {The Personality Illusion: Revealing Dissociation Between Self-Reports \& Behavior in LLMs},
author = {Pengrui Han and Rafal Kocielnik and Peiyang Song and Ramit Debnath and Dean Mobbs and Anima Anandkumar and R. Michael Alvarez},
year = {2025},
eprint = {2509.03730},
archivePrefix = {arXiv}
}

@misc{jiang2024personallm,
title = {PersonaLLM: Investigating the Ability of Large Language Models to Express Personality Traits},
author = {Hang Jiang and Xiajie Zhang and Xubo Cao and Cynthia Breazeal and Deb Roy and Jad Kabbara},
year = {2024},
eprint = {2305.02547},
archivePrefix = {arXiv}
}

@inproceedings{li2025big5chat,
title = "{BIG}5-{CHAT}: Shaping {LLM} Personalities Through Training on Human-Grounded Data",
author = "Li, Wenkai  and
  Liu, Jiarui  and
  Liu, Andy  and
  Zhou, Xuhui  and
  Diab, Mona T.  and
  Sap, Maarten",
booktitle = "Proceedings of the 63rd Annual Meeting of the {A}ssociation for {C}omputational {L}inguistics (Volume 1: Long Papers)",
month = jul,
year = "2025",
address = "Vienna, Austria",
publisher = "Association for Computational Linguistics",
url = "https://aclanthology.org/2025.acl-long.999/",
doi = "10.18653/v1/2025.acl-long.999",
pages = "20434--20471",
ISBN = "979-8-89176-251-0",
}

@ARTICLE{kuhail2024assessing,
author = {Kuhail, Mohammad Amin and Bahja, Mohamed and Al-Shamaileh, Ons and Thomas, Justin and Alkazemi, Amina and Negreiros, Joao},
journal = {IEEE Access},
title = {Assessing the Impact of Chatbot-Human Personality Congruence on User Behavior: A Chatbot-Based Advising System Case},
year = {2024},
volume = {12},
number = {},
pages = {71761-71782},
doi = {10.1109/ACCESS.2024.3402977}}

@misc{sonlu2024effects,
title = {The Effects of Embodiment and Personality Expression on Learning in LLM-based Educational Agents},
author = {Sinan Sonlu and Bennie Bendiksen and Funda Durupinar and Uğur Güdükbay},
year = {2024},
eprint = {2407.10993},
archivePrefix = {arXiv}
}

@inproceedings{kovacevic2024chatbots,
author = {Kova\v{c}evi\'{c}, Nikola and Boschung, Tobias and Holz, Christian and Gross, Markus and Wampfler, Rafael},
title = {Chatbots With Attitude: Enhancing Chatbot Interactions Through Dynamic Personality Infusion},
year = {2024},
isbn = {9798400705113},
publisher = {Association for Computing Machinery},
address = {New York, NY, USA},
url = {https://doi.org/10.1145/3640794.3665543},
doi = {10.1145/3640794.3665543},
booktitle = {Proceedings of the 6th ACM Conference on Conversational User Interfaces},
articleno = {23},
numpages = {16},
location = {Luxembourg, Luxembourg},
series = {CUI '24}
}

@inproceedings{moilanen2022effect,
author = {Moilanen, Joonas and Visuri, Aku and Suryanarayana, Sharadhi Alape and Alorwu, Andy and Yatani, Koji and Hosio, Simo},
title = {Measuring the Effect of Mental Health Chatbot Personality on User Engagement},
year = {2022},
isbn = {9781450398206},
publisher = {Association for Computing Machinery},
address = {New York, NY, USA},
url = {https://doi.org/10.1145/3568444.3568464},
doi = {10.1145/3568444.3568464},
booktitle = {Proceedings of the 21st International Conference on Mobile and Ubiquitous Multimedia},
pages = {138–150},
numpages = {13},
location = {Lisbon, Portugal},
series = {MUM '22}
}

@inproceedings{lee2024chatfive,
author = {Lee, Jungjae and Choi, Yubin and Song, Minhyuk and Park, Sanghyun},
title = {ChatFive: Enhancing User Experience in Likert Scale Personality Test through Interactive Conversation with LLM Agents},
year = {2024},
isbn = {9798400705113},
publisher = {Association for Computing Machinery},
address = {New York, NY, USA},
url = {https://doi.org/10.1145/3640794.3665572},
doi = {10.1145/3640794.3665572},
booktitle = {Proceedings of the 6th ACM Conference on Conversational User Interfaces},
articleno = {36},
numpages = {8},
location = {Luxembourg, Luxembourg},
series = {CUI '24}
}

@misc{soderqvist2025personality,
author = {S{\"o}derqvist, Emil},
institution = {Linköping University, Department of Computer and Information Science},
pages = {39},
school = {Linköping University, Department of Computer and Information Science},
title = {Personality-Matched AI Chatbots: Measuring User Experience Based on Extraversion Scores},
year = {2025}
}

@inproceedings{karnam2026bowling,
author = {Karnam, Sai Keerthana and Dash, Abhisek and Gummadi, Krishna and Mukherjee, Animesh and Weber, Ingmar and Zannettou, Savvas},
title = {Bowling with ChatGPT: On the Evolving User Interactions with Conversational AI Systems},
year = {2026},
isbn = {9798400723070},
publisher = {Association for Computing Machinery},
address = {New York, NY, USA},
url = {https://doi.org/10.1145/3774904.3792978},
doi = {10.1145/3774904.3792978},
booktitle = {Proceedings of the ACM Web Conference 2026},
pages = {9711–9721},
numpages = {11},
location = {Dubai, United Arab Emirates},
series = {WWW '26}
}

@article{nass2000machines,
  title={Machines and mindlessness: Social responses to computers},
  author={Nass, Clifford and Moon, Youngme},
  journal={Journal of social issues},
  volume={56},
  number={1},
  pages={81--103},
  year={2000},
  publisher={Wiley Online Library}
}

@inproceedings{nass1993anthropomorphism,
  title={Anthropomorphism, agency, and ethopoeia: computers as social actors},
  author={Nass, Clifford and Steuer, Jonathan and Tauber, Ellen and Reeder, Heidi},
  booktitle={INTERACT'93 and CHI'93 conference companion on Human factors in computing systems},
  pages={111--112},
  year={1993}
}

@article{peter2025benefits,
  title={The benefits and dangers of anthropomorphic conversational agents},
  author={Peter, Sandra and Riemer, Kai and West, Jevin D},
  journal={Proceedings of the National Academy of Sciences},
  volume={122},
  number={22},
  pages={e2415898122},
  year={2025},
  publisher={National Academy of Sciences}
}

@incollection{fogg2003computers,
title = {Chapter 5 - Computers as persuasive social actors},
booktitle = {Persuasive Technology},
publisher = {Morgan Kaufmann},
address = {San Francisco},
pages = {89-120},
year = {2003},
series = {Interactive Technologies},
isbn = {978-1-55860-643-2},
doi = {https://doi.org/10.1016/B978-155860643-2/50007-X},
url = {https://www.sciencedirect.com/science/article/pii/B978155860643250007X},
author = {B.J. Fogg},
}

@inproceedings{li2024evalsafety,
    title = "Evaluating Psychological Safety of Large Language Models",
    author = "Li, Xingxuan  and
      Li, Yutong  and
      Qiu, Lin  and
      Joty, Shafiq  and
      Bing, Lidong",
    booktitle = "Proceedings of the 2024 Conference on Empirical Methods in Natural Language Processing",
    month = nov,
    year = "2024",
    address = "Miami, Florida, USA",
    publisher = "Association for Computational Linguistics",
    url = "https://aclanthology.org/2024.emnlp-main.108/",
    doi = "10.18653/v1/2024.emnlp-main.108",
    pages = "1826--1843",
}

@misc{python2024,
  author = {{Python Software Foundation}},
  title = {Python Language Reference},
  version = {3},
  year = {2024},
  url = {http://www.python.org}
}

@techreport{vanrossum1995python,
  author = {van Rossum, Guido},
  title = {Python Tutorial},
  number = {CS-R9526},
  institution = {Centrum voor Wiskunde en Informatica (CWI)},
  address = {Amsterdam},
  month = {May},
  year = {1995}
}

@incollection{john1999bfi,
  author    = {John, Oliver P. and Srivastava, Sanjay},
  title     = {The Big-Five trait taxonomy: History, measurement, and theoretical perspectives},
  booktitle = {Handbook of Personality: Theory and Research},
  editor    = {Pervin, Lawrence A. and John, Oliver P.},
  volume    = {2},
  pages     = {102--138},
  year      = {1999},
  publisher = {Guilford Press},
  address   = {New York}
}

@article{lynn1986determination,
  title={Determination and quantification of content validity},
  author={Lynn, Mary R},
  journal={Nursing research},
  volume={35},
  number={6},
  pages={382--386},
  year={1986},
  publisher={LWW}
}

@article{polit2007CVI,
	title = {Is the {CVI} an acceptable indicator of content validity? {Appraisal} and recommendations},
	volume = {30},
	copyright = {http://onlinelibrary.wiley.com/termsAndConditions\#vor},
	issn = {0160-6891, 1098-240X},
	shorttitle = {Is the {CVI} an acceptable indicator of content validity?},
	url = {https://onlinelibrary.wiley.com/doi/10.1002/nur.20199},
	doi = {10.1002/nur.20199},
	language = {en},
	number = {4},
	urldate = {2026-05-15},
	journal = {Research in Nursing \& Health},
	author = {Polit, Denise F. and Beck, Cheryl Tatano and Owen, Steven V.},
	month = aug,
	year = {2007},
	pages = {459--467},
}

@book{gwet2014interrater,
  author    = {Gwet, Kilem Li},
  title     = {Handbook of Inter-Rater Reliability: The Definitive Guide to Measuring the Extent of Agreement Among Raters},
  edition   = {4},
  year      = {2014},
  publisher = {Advanced Analytics, LLC},
  address   = {Gaithersburg, MD},
  isbn      = {978-0-9708062-8-4}
}

@article{landis1977observeragreement,
 ISSN = {0006341X, 15410420},
 URL = {http://www.jstor.org/stable/2529310},
 author = {J. Richard Landis and Gary G. Koch},
 journal = {Biometrics},
 number = {1},
 pages = {159--174},
 publisher = {International Biometric Society},
 title = {The Measurement of Observer Agreement for Categorical Data},
 urldate = {2026-05-15},
 volume = {33},
 year = {1977}
}

@book{schumacker2004beginners,
  author = {Schumacker, R. E. and Lomax, R. G.},
  title = {A beginner's guide to structural equation modeling, Second edition},
  publisher = {Lawrence Erlbaum Associates},
  address = {Mahwah, NJ},
  year = {2004}
}

@Manual{r2021,
title = {R: A Language and Environment for Statistical Computing},
author = {{R Core Team}},
organization = {R Foundation for Statistical Computing},
address = {Vienna, Austria},
year = {2021},
url = {https://www.R-project.org/},
}

@manual{mcconochie2007bfi,
  author       = {William A. McConochie},
  title        = {The Big Five Inventory (BFI) Manual},
  organization = {TestMaster, Inc.},
  address      = {Eugene, OR},
  year         = {2007},
  note         = {Version 2-19-07}
}

@article{johnson2014ipipneo,
title = {Measuring thirty facets of the Five Factor Model with a 120-item public domain inventory: Development of the IPIP-NEO-120},
journal = {Journal of Research in Personality},
volume = {51},
pages = {78-89},
year = {2014},
issn = {0092-6566},
doi = {https://doi.org/10.1016/j.jrp.2014.05.003},
url = {https://www.sciencedirect.com/science/article/pii/S0092656614000506},
author = {John A. Johnson},
}

@book{gorsuch1983factor,
  author    = {Richard L. Gorsuch},
  title     = {Factor Analysis},
  edition   = {2},
  year      = {1983},
  publisher = {Lawrence Erlbaum Associates},
  address   = {Hillsdale, NJ}
}

@book{george2003spss,
  author    = {Darren George and Paul Mallery},
  title     = {SPSS for Windows Step by Step: A Simple Guide and Reference, 11.0 Update},
  edition   = {4},
  year      = {2003},
  publisher = {Allyn and Bacon},
  address   = {Boston, MA},
  isbn      = {0205375529}
}

@misc{vrabcova2025negation,
      title={Negation: A Pink Elephant in the Large Language Models' Room?}, 
      author={Tereza Vrabcová and Marek Kadlčík and Petr Sojka and Michal Štefánik and Michal Spiegel},
      year={2025},
      eprint={2503.22395},
      archivePrefix={arXiv},
      primaryClass={cs.CL},
      url={https://arxiv.org/abs/2503.22395}, 
}

@article{peereboom2025cognitive,
title = {Cognitive phantoms in large language models through the lens of latent variables},
journal = {Computers in Human Behavior: Artificial Humans},
volume = {4},
pages = {100161},
year = {2025},
issn = {2949-8821},
doi = {https://doi.org/10.1016/j.chbah.2025.100161},
url = {https://www.sciencedirect.com/science/article/pii/S2949882125000453},
author = {Sanne Peereboom and Inga Schwabe and Bennett Kleinberg},
}

\end{document}